\documentstyle[12pt]{article}
\begin{document}
\title {\bf Strange matter and its stability in presence of magnetic field}
\author{Pradip K Sahu\\
Theory Group, Physical Research Laboratory, Ahmedabad
380 009, India;\\ E-mail: pradip@prl.ernet.in.}
\maketitle
\begin{abstract}
We study the effect of a magnetic field on the interacting quark
matter and apply to strange star. We consider the low temperature
approximation to strange matter. We find that the interacting strange
quark matter is more stable compare to free quark gas in presence of
strong external magnetic field with zero and finite temperature. We
then calculate strange star structure parameters such as mass and
radius and find that the strange star is more compact for interacting
quark matter than free quark matter in presence of strong magnetic field.
\end{abstract}
\vskip 0.4in
\noindent {\it Subject Headings}: dense matter- elementary particles - stars:
neutron - stars: quark- stars: magnetic field: general
\vskip 0.4in
It has been argued that there is a strong magnetic field in the vicinity
of astrophysical objects such as neutron stars, white dwarfs and supernovae.
Particularly, the magnetic field is very high in the surface of the
neutron stars. The estimation of the magnetic field strength at the surface
of neutron stars is done in theoretical models of pulsar emission,
(Ruderman 1972) the accretion flow in the binary X-ray sources
(Ghosh and Lamb 1978) and observation of features in the spectra of
pulsating X-ray sources e.g., cyclotron lines (Trumper,
J. et al. 1978, Wheaton et al. 1979, Gruber et al. 1980 and
Mihara et al. 1990). It has been observed from a sample of more than
400 pulsars, the surface magnetic field strength is in the range
$2\times 10^{10}$G $\le H\le 2\times 10^{13}$G.
\par
Very recently, two different physical mechanism leading to an
amplification of some initial magnetic field in a collapsing star
have been proposed by several authors (Duncan and Thompson 1992,
Thompson and Duncan 1993 and Bisnovatyi- Kogan 1993).
In the first scenario (Bisnovatyi -Kogan 1993), the field amplification
leads to the formation of an additional toroidal field from the
poloidal one by twisting of the field lines due to rapid differential
rotation of proto- neutron stars. The induced toroidal magnetic field
could be as large as $H\sim 10^{15}-10^{17}$G, after the first twenty
seconds of the life of a new-born neutron star. The second mechanism
(Duncan and Thompson 1992, Thompson and Duncan 1993) is a dynamo action
in a differentially rotating and convective young neutron star, which
is responsible for the strengthenning of some initial dipole field
upto values of $H\sim 10^{12}-3\times 10^{13}$G or to $H\sim 10^{14}-
10^{15}$G. The former value is due to the convective episodes arose
during the main- sequence and later value is so, if the dipole field
is generated after the collapse. So, the new born neutron stars have
values of magnetic field as strong as $H\sim 10^{14}-10^{16}$G, or even
more.  In the interior
of neutron star, it probably reaches $\sim 10^{18}$G. Therefore, it is
advisable to study the effect of strong magnetic field and
corresponding quantum corrections on compact neutron stars.
\par
There are strong reasons for believing that the hadrons are composed of quarks
, and the idea of quark stars has already been existed for about twenty years.
If the neutron matter density at the core of neutron stars exceeds a few times
normal nuclear density a deconfining phase
transition to quark matter may take place. As a consequence, a normal neutron
star will be converted to a hybrid star with an infinite cluster of quark
matter core and a crust of neutron matter. In 1984, Witten suggested that
strange matter, e.g., quark matter with strangeness per baryon of order unity,
may be the true ground state (Witten 1984). The properties of
strange matter at zero
pressure and zero temperature were subsequently examined and it was found
that the strange matter can indeed be stable for a wide range of parameters in
the strong interaction calculations (Farhi and Jaffe 1984).
Therefore, at the core,
the strange
quarks will be produced through the weak decays of light quarks (u and d
quarks) and ultimately a chemical equilibrium will be established among the
participants. Since, the strange matter is energetically
favorable over neutron
matter, there is a possibility that whole star may be converted to a strange
star.
\par
In the present paper, we study the effect of strong magnetic
field on strange quark matter including interaction upto one
order of strong coupling constant. Throughout the calculation,
as a first
approximation, we have assumed the existence of constant uniform magnetic
field $H\approx 10^{16}$ G in the strange
star. The interior of the quark star has high conductivity
and has therefore zero electric field and uniform current density. The
magnetic field $H~\propto~ \frac{I}{R^3}~ \propto ~J~\sim ~constant$,
where $I$ is the total current, $J$ is current density and $R$ is the
radius of the spherical quark star. Taking into account the low -
temperature corrections in the thermodynamic potentials, we
investigate the stability of strange matter for finite
temperature, finite chemical potentials and finite magnetic
field. At the end, we apply strange quark matter equation of state to
structure equations of a relativistic spherical static star to calculate
the structure parameters of strange quark stars.
\par
For a constant magnetic field along the z-axis ($\vec A =(Hy,0,0))$,
the single energy eigenvalue is given by (Landau
and Lifshitz 1965)
\begin{equation}
\varepsilon_{p,n,s}=\sqrt{{p_i}^2+{m_i}^2+q_iH(2n+s+1)}~,
\end{equation}
where n=0, 1, 2, ..., being the principal quantum numbers for allowed Landau
levels, $s=\pm1$ refers to spin up(+) and down($-$) and $p_i$ is the component
of particle(species i) momentum along the field direction. Setting
$2n+s+1=2\nu$, where
$\nu = 0, 1, 2...,$ we can rewrite the single particle energy eigenvalue
in the following form
\begin{equation}
{\varepsilon_{i}}=\sqrt{{p_i}^2+{m_i}^{2}+2\nu q_iH}~.
\end{equation}
Now, it is very easy to see that $\nu=0$ state is singly degenerate,
whereas, all other states with $\nu\not=0$ are doubly degenerate.
Thus we set $b_{\nu}=2-\delta_{\nu 0}$. For $\nu=0$, $b_{\nu}=1$, which
is the lowest Landau level.
Since the temperature $T<<\mu_i$ at the core of quark star, the presence
of anti-particles can be ignored. Now instead of infinity the upper limit
of $\nu$ sum can be obtained from the following relation
\begin{equation}
{p_{Fi}}^{2}={\mu_i}^2-{m_i}^2-2\nu q_iH \ge 0,
\end{equation}
where $p_{Fi}$ is the Fermi momentum of the species $i$, which gives
\begin{equation}
\nu \le \frac{{\mu_i}^2-{m_i}^2}{2q_iH}={\nu_{max}}^{(i)}~~(nearest~~integer).
\end{equation}
Therefore, the upper limit is not necessarily same for all the components.
As is well known, the energy of a charged particle changes
significantly in the quantum limit if the magnetic field strength
is equal to or greater than some critical value $H^{(c)} =
{m_i}^2c^3/(q_i\hbar)$ (in G), where $m_i$ and $q_i$ are
respectively the mass and the absolute value of charge of
particle $i$, $\hbar$ and $c$ are the reduced Planck constant
and velocity of light respectively, both of which along with
Boltzman constant $k_B$ are taken to be unity in our choice of
units. For an electron of mass 0.5 MeV, this critical field as
mentioned above is $\sim 4.4\times 10^{13}$G, whereas for a
light quark of current mass 5 MeV, this particular value becomes
$\sim 4.4\times 10^{15}$G.
\par
Then the thermodynamic potential in presence of strong magnetic
field $H(>H^{(c)}$, critical value discussed later) is given by
\begin{equation}
\Omega_i=-\frac{g_iq_iHT}{4\pi^2}\int d\varepsilon_i\sum_{\nu}
^{\nu_{max}} b_{\nu} \frac{dp_i}
{d\varepsilon_i}\ln [1+exp(\mu_i-\varepsilon_i)/T],
\end{equation}
\noindent where $g_i$ is the degeneracy of the species $i$.
\par
Integrating by parts and substituting
\begin{equation}
p_{i}=\pm \sqrt{\varepsilon_i^2-m_i^2-2\nu q_i H},\label{eq:en}
\end{equation}
for all T, one finds
\begin{equation}
\Omega_i=-\frac{g_iq_iH}{4\pi^2}\int
d\varepsilon_i\sum_{\nu}^{\nu_{max}} b_{\nu} \frac{2\sqrt{
\varepsilon_i^2-m_i^2-2\nu q_iH}}{[exp(\varepsilon_i-\mu_i)/T+1]}
\end{equation}
where the sum over $\nu$ is restricted by the condition $\varepsilon>
\sqrt{m^2+2\nu q H}$ and the factor 2 takes into account the freedom
of taking either sign in eq(\ref{eq:en}).
For T =0, therefore, approximate the Fermi distribution by a step function
and interchange the order of the summation over $\nu$ and integration
over $\varepsilon$,
\begin{eqnarray}
{\Omega_i}^1&=&-\frac{g_iq_iH}{2\pi^2}\sum_{\nu}^{\nu_{max}}
b_{\nu} \int_{\sqrt{m_i^2+2\nu q_i H}}^{\mu}
d\varepsilon_i\sqrt{ \varepsilon_i^2-m_i^2-2\nu q_iH} \nonumber\\
&=&-\frac{g_iq_iH}{4\pi^2}\sum_{\nu}^{\nu_{max}}
b_{\nu}
\left\{
\mu_i\sqrt{\mu_i^2-m_i^2-2\nu q_i H}
\right.\nonumber\\
        && \left.
- (m_i^2+2\nu q_i H) \ln \frac{\mu_i+\sqrt{\mu_i^2-m_i^2-2\nu q_i H}}
{\sqrt{m_i^2 +2\nu q_i H}}
\right\} ,
\label {eq:om}
\end{eqnarray}
\noindent
which is first order thermodynamic potential.
\par
Considering interaction upto one order of strong coupling
constant $\alpha_c$ (Freedman and McLerran 1978, Alcock et al.
1986 and Haensel et al. 1986), the second order thermodynamic
potential in presence of strong magnetic field for zero
temperature is given by
\begin{equation}
{\Omega_i}^2 =
-\frac{g_iq_iH}{4\pi^2}\frac{8\alpha_c}{\pi} \sum_{\nu}^{\nu_{max}}
b_{\nu}
\left\{
(\mu_i^2 - m_i^2 - 2\nu q_i H) \ln \frac{{\mu_i}^2
- {m_i}^2 - 2\nu q_i H}{q_i H}
\right\}
\end{equation}
\noindent The low - temperature corrections to the first order
and second order thermodynamic potentials have been derived in
Ref. (Sahu 1995) and these are:

\begin{eqnarray}
       {\Omega_i}^1 =\nonumber\\
        &&  -\frac{g_iq_i H}{4\pi^2}
        \sum_{\nu=0}^{ \nu_{max}}
        b_\nu
\left\{
\mu_i \sqrt{\mu_i^2-{m_i}^2-2q_i H\nu}
        - ({m_i}^2+2q_iH\nu)
        \ln
        \frac{ \mu_i + \sqrt{\mu_i^2-{m_i}^2-2q_iH\nu}}
{\sqrt{{m_i}^2+2q_iH\nu}}
\right.\nonumber\\
        & &\nonumber\\
        && \left.
+ \frac{T^2\pi^2}{6} \frac{\mu_i}{(\mu_i^2 - {m_i}^2 - 2q_iH\nu)^{1/2}}
\right\},
\end{eqnarray}

\begin{eqnarray}
       {\Omega_i}^2 =\nonumber\\
        &&  -\frac{g_iq_i H}{4\pi^2}
        \sum_{\nu=0}^{ \nu_{max}}
        b_\nu
\frac{8\alpha_c}{\pi}
\left\{
(\mu_i^2 -{m_i}^2 -2q_iH\nu)
\ln \frac{\mu_i^2 -{m_i}^2 -2q_iH\nu}{q_iH}
\right. \nonumber\\
& &\nonumber\\
&& \left.
    + \frac{T^2\pi^2}{6}
\left[
1 + \frac{2\mu_i^2}{\mu_i^2 - {m_i}^2
- 2q_iH\nu} + \ln \frac{\mu_i^2 -{m_i}^2 -2q_iH\nu}{q_iH}
\right]
\right\}.
\end{eqnarray}
\par
This expansion is valid for $ \frac{T}{\mu_i - \sqrt{m_i^2 + 2q_iH\nu}}
\ll 1$. This means that the distance from the edge of any Landau level
$\varepsilon_{\nu}(p_{z}) = \sqrt{m_i^2 + 2q_iH\nu}$ to the Fermi
surface $\mu_i$ is
much greater than the temperature, or in otherwords, the Landau level
with $\nu=\nu_{max}$ should be partially filled, i.e., its edge cannot
concide with the Fermi surface.
\par
So, the total thermodynamic potential is

\begin{equation}
{\Omega_i}(T,H,\mu_i)=
{\Omega_i}^1(T,H,\mu_i)+ {\Omega_i}^2(T,H,\mu_i).
\label{total}
\end{equation}
\par
In our study, we assume that strange quark matter is charge
neutral and also chemical equilibrium, then
\begin{equation}
\mu_d=\mu_s=\mu=\mu_u+\mu_e, \label{eq:ch}
\end{equation}
and charge neutrality conditions gives
\begin{equation}
2n_u-n_d-n_s-3n_e=0. \label{eq:cha}
\end{equation}
The baryon number density of the system is given by
\begin{equation}
n_B=\frac{1}{3}(n_u+n_d+n_s) \label{eq:ba}.
\end{equation}
Using above eqs(\ref{eq:ch}, \ref{eq:cha}, \ref{eq:ba}) one can solve
numerically for the chemical potentials of all the flavors and
electron. Having the expression for the total thermodynamic
potential, we may move forward to calculate the number density
($-\partial \Omega_i/\partial \mu_i$) and the magnetization
($-\partial \Omega_i/\partial H$) of the species $i$
($u,~d,~s,~e$) (Sahu 1995). Using Eq. (\ref{total}) and definition of number
density,
one has the following expression:
        \begin{eqnarray}
        \lefteqn{n_i(H,\mu,T) =}\nonumber\\
&& \frac{g_iq_iH}{2\pi^2} \sum_{\nu=0}^{\left[  \frac{\mu_i^2 -
m_i^2}{2q_iH} \right]}
b_{\nu} \sqrt{\mu_i^2-m_i^2-2q_iH\nu}
\left\{
1 - \frac{T^2\pi^2}{12} \frac{m_i^2 +
        2q_iH\nu}{(\mu_i^2-m_i^2-2q_iH\nu)^2}
\right.\nonumber\\
& &\nonumber\\
&& \left.
+\frac{8 \alpha_c}{\pi}
\left[
\frac{\mu_i}{\sqrt{\mu_i^2 -m_i^2
-2q_iH\nu}}
\left(
1 + \ln \frac{\mu_i^2 -m_i^2 -2q_iH\nu}{q_iH}
\right)
\right. \right.\nonumber\\
& &\nonumber\\
&& \left. \left.
-\frac{T^2\pi^2}{6}
\mu_i
\left(
\frac{3(m_i^2+2q_iH\nu)-\mu_i^2}{{(\mu_i^2-m_i^2-2q_iH\nu)}^{5/2}}
\right)
\right]
\right\}
\label{den}
       \end{eqnarray}
\noindent
Now, for $T=0$, we have the number density of the species $i$:

        \begin{eqnarray}
n_i(H,\mu,T=0) =\nonumber\\
 \frac{g_iq_iH}{2\pi^2}
\sum_{\nu=0}^{\left[  \frac{\mu_i^2 -
m_i^2}{2q_iH} \right]}
b_{\nu}
\left\{
\sqrt{\mu_i^2-m_i^2-2q_iH\nu}
+
\frac{8 \alpha_c}{\pi}
\left[
\mu_i
\left(
1 + \ln \frac{\mu_i^2 -m_i^2 -2q_iH\nu}{q_iH}
\right)
\right]
\right\}
\label{den0}
    \end{eqnarray}
\par
In figure 1, we have plotted the number density of $u$ quarks as
a function of magnetic field for fixed chemical potential ($\mu_u=20$ MeV).
The $u$ quark density is showing an oscillating behavior as consecutive
Landau levels are passing the Fermi level. The number density is low for
case (a), where $\alpha_c=0.0$ and the number density increases with increase
in $\alpha_c=0.01$ for case (b) and $\alpha_c=0.05$ for case (c)
respectively. Thus, we noticed that the number density is high
for strongly interacting quarks.
\par
In the weak magnetic field limit $(H_0\ll (\mu_i^2 - m_i^2))$, one may
reduce the number density as follows:
 \begin{eqnarray}
        \lefteqn{n_i(H_0\ll(\mu_i^2-m_i^2),\mu,T=0) }
\nonumber\\
&&  \approx \frac{g_i}{4\pi^2}
\left\{
\frac{2}{3}(\mu_i^2-m_i^2)^{3/2}
\right.\nonumber\\
&& \left.
+\frac{8\alpha_c}{\pi}
\left[
\mu_i (\mu_i^2 -m_i^2)\ln\| \frac{\mu_i^2 -m_i^2}{q_iH_0}\|
\right]
\right\}
\label{denH0}
       \end{eqnarray}
\noindent
where, we can substitute in the limit $2q_iH\nu = \theta$
\begin{equation}
\sum_{\nu=0}^{\left[ \frac{\mu_i^2 -m_i^2}{2q_iH} \right]}
= \int_{0}^{\left[ \frac{\mu_i^2 -m_i^2}{2q_iH} \right]} d\!~~ \nu
= \lim_{H\rightarrow 0} \frac{1}{2q_iH} {\int_{0}^{(\mu_i^2-m_i^2)}} d
\!~~ \theta
\end{equation}
\noindent
The ratio of $u$ quark density, $n_u(H_0\ll(\mu_u^2-m_u^2))/n_u(H)$ as
function of chemical potentials for fixed magnetic field, curve (a) $H\approx
5\times 10^{15}$ G and curve (b) $H\approx 10^{16}$ G, is shown
in figure 2. We choose the weak field limit, $H_0$ to be $4.4\times 10^{13}$ G.
As the chemical potential increase, the ratio decreases and reaches to unity
for
both curves (a) and (b). We noticed that the magnetic field has significant
contribution to the quark densities and hence to the total thermodynamic
potential. For illustration purpose, we choose here the $u$ quark density.
\par
{}From the definition of magnetization
$M=\displaystyle-\frac{\partial {\Omega_i}}{\partial H}$, where
$M(H,\mu,T) = M(H) + \tilde{M}(H,\mu,T)$, $M(H)$ is the vacuum
magnetization, one has

        \begin{eqnarray}
        \lefteqn{\tilde{M}(H,\mu,T=0)=}\nonumber\\
        &&
        \frac{g_iq_i}{4\pi^2}
        \sum_{\nu=0}^{\left[  \frac{\mu_i^2 - m_i^2}{2q_iH} \right]}
        b_{\nu}
        \left\{
        \mu_i    \sqrt{\mu_i^2-m_i^2-2q_iH\nu}
         - (m_i^2+4q_iH\nu) \ln
        \left( \frac{ \mu_i +
        \sqrt{\mu_i^2-m_i^2-2q_iH\nu}}
        {\sqrt{m_i^2+2q_iH\nu}}
                        \right)
\right.\nonumber\\
        &&\nonumber\\
        &&
        \left.
    +\frac{8 \alpha_c}{\pi}
\left[
(\mu_i^2 -m_i^2 -4q_iH\nu) \ln
\frac{\mu_i^2 -m_i^2 -2q_iH\nu}{q_iH} -(\mu_i^2 -m_i^2)
\right]
\right\}.
        \end{eqnarray}

\noindent
One may calculate the magnetic susceptibility $\chi$, which  is
defined as $(\partial M/\partial H)$.
\par
Figure 3 shows the magnetization as a function of magnetic field
for fixed chemical potential ($\mu_u =20$ MeV). We have not
considered the vacuum magnetization in this figure, because the
vacuum contribution is small (Sahu 1995). We noticed that the
quark gas exhibits the de Hass - van Alphen effect. We considered
the strong coupling constant in curves (a) $\alpha_c=0.0$, (b) $\alpha_c=0.01$
and (c) $\alpha_c=0.05$ respectively for illustration purpose. The
magnetization
is high at small value magnetic field and approaches to case (a) (e.g., free
quark gas) for large magnetic field. Thus the de Hass - van Alphen effect
amplifies with increase in strength of the strong coupling constant.
\par
The total energy density and the external pressure of the
strange quark matter is given respectively by
\begin{eqnarray}
\varepsilon &=& \sum_{i}\Omega_i +B +\sum_{i}n_i \mu_i -
T(\frac{\partial \Omega_i}{\partial T})_{\mu_i} \nonumber \\
p&=&-\sum_i\Omega_i-B,
\end{eqnarray}
where $i=~u,~d,~s,~e$.
\noindent
The last term is the total energy density is entropy of the
system, which is non zero and is given by

\begin{eqnarray}
       -T(\frac{\partial\Omega_i}{\partial T})_{\mu_i}
\nonumber\\
        && = \frac{g_iq_i H}{4\pi^2}
        \sum_{\nu=0}^{ \left[   \frac{\mu_i^2 - {m_i}^2}{2q_iH}   \right]}
        b_\nu
\frac{T^2\pi^2}{3}
\left\{
        \frac{\mu_i}{(\mu_i^2 - {m_i}^2 - 2q_iH\nu)^{1/2}}
\right.\nonumber\\
        & &\nonumber\\
        && \left.
+\frac{8\alpha_c}{\pi}
\left[
1 + \frac{2\mu_i^2}{\mu_i^2 - {m_i}^2
- 2q_iH\nu} + \ln \frac{\mu_i^2 -{m_i}^2 -2q_iH\nu}{q_iH}
\right]
\right\}.
\end{eqnarray}
\noindent
Here, we have considered the conventional
bag model for the sake of simplicity in presence of magnetic field.
We are assuming that the self interacting
quarks are moving within the system and as usual the
current masses of both u and d quarks are extremely small, e.g.,
5 MeV each, whereas, for s-quark the current quark mass is to be
taken 150 MeV. We choose the bag pressure $B$ to be 56
$MeV~fm^{-3}$ and the strong coupling constant $\alpha_c < 1$.
Also, we set the magnetic field to be $H\approx10^{16}$ G in
our calculations. Since we choose the magnetic field along the z-axis,
it follow from energy-stress tensor that the constant uniform magnetic
field ($H=10^{16}$G) contribute to the pressure and the
energy density by $\frac{H^2}{8\pi}$, which is much smaller than
the bag pressure, e.g., $0.0025~<< 56 MeV~fm^{-3}$.
\par
The variation of energy per baryon with baryon number density
is shown in figure 4. The curve (a) is for free quark gas,
$\alpha_c=0.0$ at zero temperature and curve (b) is for interacting
quark gas with interaction strength $\alpha_c=0.1$ at zero temperature,
whereas, the curve (c) is for $\alpha_c=0.1$ and temperature $T=50$ MeV.
For all the cases, the magnetic field strength is $\approx 10^{16}$ G. We
concluded from the figure that the interaction strength and finite
temperature make the strange quark matter energetically more stable
compare to the zero temperature and zero interaction strength.
\par
Next, we have shown the variation of pressure with energy density in
figure 5. These are the equation of states of strange quark matter.
The curve (a) is for free quark gas ($\alpha_c=0.$) at zero temperature
and curve (b) is for interacting quark gas ($\alpha_c=0.1$) at zero
temperature. From this figure, we noticed that the interaction strength
makes the quark equation of state soft. Also, we have seen that the
softness increases with increase in interaction strength.
\par
The mass and radius for nonrotating strange quark stars are
obtained by integrating the structure equations of a
relativistic spherical static star composed of a perfect fluid
which is derived from Einstein equation. These equations are
given in Ref. (Datta et al. 1992, Sahu, Basu and Datta 1993 and
Ghosh and Sahu 1993), hence, we are not reproducing here. For a
given equation of state, and given
central density, the structure equations are integrated
numerically with the boundary conditions $m(r=0)=0$, to give $R$
and $M$. The radius $R$ is defined by the point where $p\simeq
0$. The total gravitational mass $M$, moment of inertia $I$,
surface red shift $z$ and the period $P_0$ corresponding to fundamental
frequency $\Omega_0$ are then given by
$M=m(R),~ I=I(R),~ z=(1-2GM/Rc^2)^{-1/2}-1$ and $P_0=\frac{2\pi}{\Omega_0}$
respectively,
where $\Omega_0=(\frac{3GM}{4R^3})^{1/2}$ (Cutler, Lindblom and
Splinter 1990).  These
are correspond to maximum mass of stable star that can have, presented
in Table 1. Figure 6 shows the variation of mass
with central density for two equation of states as illustrated
in Fig. 5. We noticed from this figure and table that with increase in
interaction strength, the maximum mass and the corresponding radius
decrease from 1.83 to 1.64 solar mass
and from 10.4 km to 9.58 km and therefore, the star becomes more compact.
Similarly, the values for surface red shift, moment of inertia and
fundamental period decrease. Also, we have noticed that with increase
in magnetic field strength, the density also increases and therefore, the star
becomes more compact (Chakrabarty and Goyal 1994, Chakrabarty and Sahu 1995 and
Chakrabarty 1995).
\par
In conclusion, we concluded that the presence of strong magnetic
field in interacting strange quark matter reduces the mass and radius of
strange star. That is the star becomes more compact. With inclusion of
interaction strength of quarks at finite temperature and at zero
temperature in presence of strong magnetic field, the strange quark
matter is more stable compare to free quark gas at zero temperature.
\vskip 0.5 in
\noindent {\bf RERERENCES}
\vskip 0.2 in
\noindent
Alcock, C. et al. 1986, Ap. J. 310, 261.\\
Bisnovatyi- Kogan, G. S. 1993, Astron. Astrophys. Transaction 3, 287.\\
Chakrabarty, S. and Goyal, A. 1994, Mod. Phys. Lett. A9, 3611.\\
Chakrabarty, S. and Sahu, P. K. 1995, hep-ph/9508251; PRL-TH/95-14.\\
Chakrabarty, S. 1995, Phys. Rev. D51, 4591 and references therein. \\
Cutler, C., Lindblom, L. and Splinter, R. J. 1990, Ap. J. 363, 603.\\
Datta, B., et al. 1992, Phys. Lett. B283, 313.\\
Duncan, R. C. and Thompson, C.  1992, Ap. J. 392, L9.\\
Farhi, E. and Jaffe, R. L. 1984, Phys. Rev. D30, 2379.\\
Freedman, B. and McLerran, L. 1978, Phys. Rev. D 17, 1109.\\
Ghosh, P. and Lamb, F. K. 1978, Ap. J. 223, L83.\\
Ghosh, S. K. and Sahu, P. K. 1993, Int. Jour. Mod. Phys. E2, 575.\\
Gruber, D. E. et al. 1980 Ap. J. 240, L127.\\
Haensel, P. et al. 1986 Astr., Astrophys. 160, 121.\\
Landau, L. D. and Lifshitz, E. M. 1965, Quantum
Mechanics, Pergamon Press, Oxford.\\
Mihara, T. et al. 1990, Nature 346, 250.\\
Manchester, R. N. and Taylor, J. H. 1981, Astron. J. 86,
1953.\\
Ruderman, M. A. 1972, Ann. Rev. Astron. Astrophys. 10, 427. \\
Sahu, P. K. 1995, hep-th/9509046, submitted and reference therein.\\
Sahu, P. K., Basu, R. and Datta, B. 1993, Ap. J. 416, 267.\\
Trumper, J. et al.  1978, Ap. J. 219, L105.\\
Thompson, C. and Duncan, R. C. 1993, Ap. J. 408, 194.\\
Witten, E. 1984, Phys. Rev. D30, 272.\\
Wheaton, W. A. et al. 1979, Nature 272, 240.\\
\vfil
\eject
\newpage
\vskip 0.5 in
\begin{table}
\caption {The radius ($R$), mass ($M$), surface red shift ($z$), moment
of inertia ($I$) and period of fundamental frequency ($P_0$) of strange
stars versus central density $\epsilon_c$ for two cases; (a):
$\alpha_c=0.0$ and (b):
$\alpha_c=0.1$. For both the cases  $H\approx 10^{16}$ G.
These values are correspond to maximum mass of stable strange star.}
\hskip 0.5 in
\begin{tabular}{ccccccccc}
\hline
\multicolumn{1}{c}{$\epsilon_c$} &
\multicolumn{1}{c}{$R$}&
\multicolumn{1}{c}{$M/M_{\odot}$}&
\multicolumn{1}{c}{$z$}&
\multicolumn{1}{c}{$I$}&
\multicolumn{1}{c}{$P_0$}&
\multicolumn{1}{c}{$Cases$}\\
\multicolumn{1}{c}{($g~cm^{-3}$)} &
\multicolumn{1}{c}{($km$)}&
\multicolumn{1}{c}{} &
\multicolumn{1}{c}{} &
\multicolumn{1}{c}{($g~cm^{-2}$)} &
\multicolumn{1}{c}{($ms$)} &
\multicolumn{1}{c}{}\\
\hline
2.0$\times 10^{15}$&10.37&1.83&0.44&1.85$\times 10^{45}$&0.49&(a)\\
2.5$\times 10^{15}$&9.58&1.64&0.42&1.38$\times 10^{45}$&0.48&(b)\\
\hline
\end{tabular}
\end{table}
\hskip 0.5 in
\vfill
\eject
\newpage
\begin{figure}
\caption {The number density of $u$ quarks as
a function of magnetic field for fixed chemical potential ($\mu_u=20$ MeV).
Where Case (a) is for $\alpha_c=0.$, case (b) is for $\alpha_c=0.01$
and case (c) is for $\alpha_c=0.05$ respectively.}
\end{figure}
\begin{figure}
\caption { The ratio of $u$ quark density, $n_u(H_0\ll(\mu_u^2-m_u^2))/n_u(H)$
as
function of chemical potentials for fixed magnetic field, curve (a) $H\approx
5\times 10^{15}$ G and curve (b) $H\approx 10^{16}$ G.
The choice of the weak field limit, $H_0$ is $4.4\times 10^{13}$ G.}
\end{figure}
\begin{figure}
\caption { The magnetization as a function of magnetic field
for fixed chemical potential ($\mu_u =20$ MeV). The curves
(a) is for $\alpha_c=0.0$, (b) is for $\alpha_c=0.01$
and (c)$\alpha_c=0.05$ respectively.}
\end{figure}
\begin{figure}
\caption { The variation of energy per baryon with baryon number density
with constant magnetic field $H\approx 10^{16}$.
The curve (a) is for free quark gas, $\alpha_c=0.0$ at zero temperature
and curve (b) is for interacting quark gas ($\alpha_c=0.1$) at zero
temperature and curve (c) is for $\alpha_c=0.1$ and temperature $T=50$ MeV.}
\end{figure}
\begin{figure}
\caption { Pressure and energy density ($\epsilon$) curves for case (a)
$\alpha_c=0.0$
and case (b) $\alpha_c=0.1$ with constant magnetic field $H\approx 10^{16}$.}
\end{figure}
\begin{figure}
\caption { Mass and central density ($\epsilon_c$) curves for two cases as
mentioned in figure 5.}
\end{figure}
\vfil
\eject
\end{document}